# The Rise of the AI Co-Pilot:
# Lessons for Design from Aviation and Beyond

Abigail Sellen
Microsoft
Cambridge, UK
asellen@microsoft.com

Eric Horvitz
Microsoft
Redmond, WA, USA
horvitz@microsoft.com

No longer confined to the boundaries of laboratories and niche applications, generative AI methods have erupted onto the global stage with relevance to almost every aspect of knowledge work and with increasing use in everyday life. From generating communications, summarizing documents and crafting literature to engineering code, translating languages, and synthesizing videos, the magnitude of their potential impact has caught even the most forward-thinking AI experts and technology visionaries off guard. It is hard to identify aspects of our lives that will not be influenced by this technological disruption. Advances in AI will likely transform everything from routine work to highly skilled professions, including high-stakes areas like medicine, law, and education [14].

As these advancements in AI lead us all to explore new ways of harnessing these capabilities, we are also developing a new language and set of metaphors for how we talk about the technology and our interactions with it. "Bots", "agents," and "co-pilots" are some of the terms that are being given new weight through their place in our lives. This is not just about new linguistic distinctions, but they are also shared metaphors, having cultural implications for how we see the technology and expect to interact with it.

Many have argued that we need to view these technologies as collaborative partners rather than competitors—as agents that work *for* us or *with* us [7, 11], with some even calling for them to be relegated completely to the realm of smart "tools" [21]. Questions about how we design AI technologies and how we configure and see our relationship to them, are therefore of critical importance.

So how do we proceed? How do we best design technologies that can automate many aspects of human activity, have some degree of agency, and offer skills that can rival and even surpass human capabilities? As Elish [8] has noted, there is a great deal at stake. As humans increasingly find themselves engaging with complex, automated systems, questions of responsibility in the case of incidents and accidents are more important than ever, especially as humans often bear the brunt of the blame despite poorly designed systems.

To address these challenges, we believe it is important to leverage decades of research on human-machine interaction and collaboration—much of which comes from the fields of Human-Computer Interaction (HCI) and Human Factors Engineering (HFE). While HCI has wrangled for decades with designs that influence our engagement with computer technology, HFE has largely taken a separate path, exploring how human operators interact with safety-critical systems such as process control in industrial plants and aviation. Taken together, there is a growing consensus that these fields have much to tell us about designs that influence the relationship between complex systems and their users or human operators [4].

As an example, the metaphor "AI co-pilot" implies a great deal about how we expect to work with AI. This metaphor suggests a technological partner that:

- Operates under human direction and oversight.
- Can communicate in a fluid, conversational manner allowing for natural and complex interaction.
- Can be assigned some degree of agency and intrinsic motivation in working toward shared goals.
- Has a broad scope for problem solving, but also has a specific set of skills that can supplement as well as complement those of the pilot.
- Can act as a back-up for the pilot, helping to monitor a situation and taking over tasks as and when needed.

The most important point is at the top of the list: *the co-pilot is not the pilot*. The co-pilot is subservient to direction from the pilot. It is the pilot who makes the critical decisions and has ultimate responsibility for flying the aircraft. And it is the pilot who oversees the co-pilot, assigning and withdrawing the co-pilot's responsibilities. While the co-pilot has some workable grounding about the task at hand, it is assumed that their role and contribution is one part of the more comprehensive and expansive understanding and awareness the pilot has when it comes to safely flying a plane.

For these reasons, we believe the term *co-pilot* is a useful metaphor for describing how AI technology is intended to act in relation to the human user or operator. And whether we are envisioning AI systems that support people in one-off decisions, or in more ongoing and dynamic interactions with people (such as in semi-autonomous vehicles or alerting systems in medicine), the concept is a rich and useful one.

At the same time, it is also useful in leveraging prior efforts and insights from HFE. The aviation and control engineering industries have a great deal to tell us about automation, including the risks and best practices for working with automated tools.

## 2 Four Lessons from Aviation

A good starting point is a short but highly influential paper written in 1983 by Lisanne Bainbridge entitled "Ironies of Automation" [3] in which she pointed out some of the crucial design considerations needed to develop systems where humans collaborate with machines capable of automating complex tasks. Bainbridge was focused on application to process control in industrial plants and aviation flight decks. However, the parallels with human-AI collaboration are clear.

Most of Bainbridge's original observations relate to the propensity for over-reliance on automation. Her work alerts us to being cautious as we design our AI systems to concerns that are well-known in the field of HFE as follows:

**The problem of vigilance.** As the degree of automation increases, humans increasingly take on the task of *vigilance*, becoming monitors of what the system is doing rather than active participants in the workflow. We know from very early vigilance studies that humans are very poor at these kinds of tasks. For example, a classic study [13] found that when people were asked to monitor visual processes for rare events, even highly motivated subjects found it difficult to maintain attention for more than about half an hour. Further, research [e.g., 23] shows that jobs which have a heavy monitoring requirement, with little in the way of direct engagement in activity, lead to high levels of stress and poor health, something which partly explains why jobs such as air traffic control are so difficult [12]. When autopilot was first introduced in aircraft, concerns that this would turn humans into system monitors even led some HFE specialists to propose that, in the flight deck, we should ask whether it should be computers monitoring the pilots rather than the other way round [24].

Turning to today's AI systems, we should similarly ask what AI applications mean for a world in which many of our jobs might increasingly require us to monitor or oversee what our intelligent systems are doing, with concomitant concerns about whether we will notice when we need to intervene and, more long term, whether our jobs will become more stressful and less satisfying as a result.

Partly as a result of the pandemic, we are much more attuned now as a society to the importance of wellbeing at work not just for personal reasons, but also for the health and productivity of organisations. As Csikszentmihalyi [5] noted in his studies, we are at our happiest when we're absorbed in tasks which not only allow us to exercise our talents but to stretch them. This calls for careful attention to how to design the human-machine partnership around the primacy of people, to make sure that humans are engaged, that they can do what they enjoy and are good at, and that they can learn and grow as a consequence.

**The takeover challenge.** Another major concern is attending to the hand-off between human and machine during the course of work. When trouble arises, humans often need to intervene, and when processes are heavily automated there are consequences for this transition when the human has been largely out of the loop. As Bainbridge highlighted [3], having to re-engage with an automatic system can be problematic because human operators have been paying attention elsewhere, meaning they have reduced situational awareness of what the requirements of the task are and what the context of the work is, undermining the human's ability to take appropriate action.

In aviation, there have been notorious examples where, firstly, pilots have not understood, paid attention to, or been properly alerted to the actions of the autopilot system. Then, when things start to spiral out of control (sometimes quite literally) the pilot is lacking in the necessary situation awareness and knowledge to step in and take the right corrective actions. A salient example is the case of China Airlines Flight 006 in 1985 where reliance on the 747's autopilot during an engine failure introduced complexities contributing to disorientation of the pilot, leading to a sudden uncontrolled 30,000 foot plunge of the aircraft toward the ocean.

The influence of automation on human awareness and vigilance can be seen in other domains too, including our role as the operators of heavy equipment in our daily lives. Studies show that new forms of automation in our cars, such as adaptive cruise control and smart lane following, can have detrimental effects. For example, automobile drivers can become complacent and less aware of hazards with the use of adaptive cruise control while driving, resulting in a negative impact on safety, such as longer response times to hazards [19, 25].

Extrapolating to AI systems highlights the possibility that too much automation may create situations in which we over rely on systems to do our work for us, with the result that we have no clear understanding of the flow of work, where and how we should intervene, and a lack of resources in the moment to enable us to effectively respond, guide, or contribute. Keeping the human in the loop is therefore crucial not just for these reasons, but also because disengagement from tasks where a person can otherwise exercise their skills can lead to deterioration in well-being, mood, and creativity [6]. Further, this points to the need for better capabilities for mutual grounding, feedback, and alerting on the part of the AI system.

**De-skilling through automation.** Longer term, the ability to take over control is of course made worse if the flight crew has suffered from the deterioration of skills due to an ongoing lack of engagement in the flow of work. Bainbridge drew attention to the irony of not only physical skills deteriorating, but also cognitive skills waning when processes are increasingly automated. However, when automation fails, it is just these skills that are crucial when the human needs to step in and take control. The aviation industry has known this for decades [24], hence the need for simulator training where faults are routinely injected into the system to ensure that not only pilots know how to manually fly a plane, but that they also know how to diagnose a problem. The deployments and reliance on aviation autopilots, can be recognized as having serious downsides both in terms of deficits of pilots with developing an implicit understanding of how a system works, and the problem-solving skills needed to critically evaluate the output of the autopilot and the state of the aircraft and to take corrective action.

Extrapolating to human-AI systems underscores the fact that we understand very little about how the offloading of different components of our workflows to these new AI systems will impact our cognitive skills for a given task or job of work. The complexity and impenetrability of today's AI systems already compromises their intelligibility for us, but if we consider that we might be less and less "hands on" as these technologies are harnessed to do more, we might fail to build effective, implicit understandings of them. Such understandings are built through our ongoing interactions with systems, helping us to learn how our actions relate to the output of the machine. And there are many other nagging, related questions too: How deeply will we understand or remember aspects of the domain we are working within if an AI system does the work? This may not matter for some kinds of tasks, but for others, we may only ever develop a shallow understanding of the subject of our work. Further, what critical skills will wither away and which new ones will we need to develop? There is little doubt that AI will shape our cognition, which in turn has important implications for education, skills training, and jobs of the future.

**Trust and automation bias**. A fourth issue concerns the predisposition for the human user or operator to critically assess or question the output of the automated system. When autopilot was first introduced into aircraft, accident analyses found that a key contributing factor was that pilots were overly reliant and over-trusting of the autopilot system. The pilots either blindly accepted its output, or failed to act unless the autopilot system advised them or alerted them to act—a phenomenon now known as "automation bias." This perception of the superiority and correctness of the machine is one lesson from aviation. Another lesson points to the perception of the superiority of the pilot in relation to more junior crew being a contributing factor in catastrophic accidents. Lessons were learned some decades ago [20] of the risks of junior crew not speaking up or questioning the actions of the captain. Perhaps the most notorious case is that of the Tenerife airport disaster in 1977, where the reluctance of junior crew to challenge the captain of the KLM aircraft was cited as a contributing factor to the deadliest accident in aviation history. Too much trust or deference, combined with the failure to question actions in the cockpit, whether by pilot or autopilot, can lead to critical incidents.

The analogy to AI systems here should be clear. We know already that people are prone to automation bias when it comes to working with AI, tending to favor and failing to question the output of the AI system [e.g., 1]. This raises key concerns about how we design systems to engender the appropriate levels of trust in the output of the machine, and furthermore, how we can provide tools and design the experience so that people critically assess the actions and recommendations of their AI partner.

## 3 What Can We Do?

Given these concerns, what can we do to ensure that our future with AI systems keeps us engaged, in charge, and in a safe and trusted relationship with our digital co-pilot? Again, the literature in both HCI and HFE offers guidance, but we also need to extend our approach: AI systems today are more complex, are more dynamic, and are being granted more agency than ever. They are capable of simulating human behavior in new ways, and of generating plausible output which may in fact be false. Their general, polymathic powers, high availability, and ease of engagement have made them pervasive in daily life. All of this implies both that we should draw as much as possible on existing design philosophy and research across fields, but also that we need to think deeply about our new priorities when we design these systems.

Accordingly, here are some of the key insights that need to be kept top of mind:

**Keep humans in the loop**: As we have described, existing research highlights the need for human engagement (whether mental or physical) rather than complete relegation to the role of overseer or monitor. As technologists become more ambitious about the degree to which AI systems can automate human activity, humans may, at the same time, start to monitor these systems less as they trust them more. But for all the reasons cited above, users need to be kept in the loop to both be and feel empowered, to implicitly learn about the system through interaction, and for well-being. More specifically:

- Be cautious about the amount of automation possible before humans are invited or required to intervene. For example, we know that continuous and ongoing interaction with an AI system can effectively and implicitly build users' mental models of the system [22].

- Consider with care when it makes more sense that AI systems monitor humans, rather than the other way round. (This is after all what spell and grammar checkers do today, supporting and monitoring human action and offering assistance in an unintrusive way.)

**Uphold the primacy of human agency and role allocation.** It is not enough that humans are engaged—we also need to design a safe and effective partnership. Studies in HCI and HFE reach a similar conclusion on this point, but changes in the capabilities, pervasiveness, and agency of AI makes action in accordance with the findings and implications more urgent. First and foremost, as in aviation, the co-pilot must cede ultimate control and final responsibility to the human user as "pilot", with the co-pilot acting in a well-defined supporting role. Designing the role of the machine also needs to take into account what people are good at versus the capabilities of machines, as well as consideration of what motivates and makes people happy. This means we need to do the following:

- Design AI features and overall software applications they are hosted within to enable and celebrate the primacy of human agency. Achieving this goal can be challenging and thus may tend to be lost to simpler, easier integrations of AI into applications.

- Reinforce a clear hierarchy of control and decision-making with the human in charge. Note that this does not mean that such oversight necessarily leads to better outcomes in all situations. In fact, research shows that predictive or diagnostic tasks may be more accurately achieved by the machine [9]. However, such reliance on human oversight makes explicit that the pilot or human is the one held accountable, with the responsibility to ensure that the best sources of information are evaluated and that the implications of output or recommended actions amidst the complexities of the open world are considered.

- Allow users control over how and when they engage in the workflow.

- Support users in carrying out aspects of tasks which allow for self-expression, creativity, social judgment, and navigating complex situations that extend beyond AI capabilities.

**Building human skills and capabilities.** AI systems are now capable of taking on more of our human tasks than ever before, changing the skills we learn and maintain. However, when designed right, AI systems can enable us to build new skills rather than to undermine our existing capabilities. For example, educators are now exploring ways in which AI can be used as personalised coaches or tutors to boost our capabilities [e.g., 16]. Consider the following:

- Though AI systems are developing sophisticated conversational capabilities, users nonetheless need to learn how to "speak machine" to best effect both in terms of learning how to most effectively prompt these systems as well as how to assess their actions.

- AI systems need to be endowed with the ability to collaborate with users to achieve mutual understanding about the intentions, goals and capabilities of users. One approach is the development of teachable AI systems (training the co-pilot) so that they are personalised and work for us in different contexts [17].

- Consider the coaching metaphor—design systems which help us develop new skills that continue to exist even when not using the AI system [10].

**Designing more intelligible systems.** By working to boost the intelligibility of AI systems, we support better situational awareness and mental models for users so that they can intervene when necessary, critically assess the output of the machine, and address problems when they occur. Tried and tested approaches from HCI are useful here, such as:

- Refer to well established HCI methods and design guidelines, including principles specifically for *mixed-initiative interfaces*, a term coined by Horvitz [11] referring to systems where people and machines work in partnership toward shared goals. More recently, the set of principles for guiding designs for human-AI interaction were further extended and studied [2].

- General HCI principles include making clear what a system is capable of, acknowledging user actions, providing effective and ongoing feedback to users, and delivering well designed explanations for system responses, and alerting users to problems or issues in intelligent ways, at the right time and with guidance as to how to intervene.

- While "explainable AI" is a burgeoning field, it is also clear that social science has a great deal to contribute here [e.g., 15] and that we need to consider how to design the whole user experience, not just the output of the algorithms [22].

**Designing for appropriate levels of trust.** The sheer complexity of today's AI models, the fact that they are probabilistic, and their tendency to produce plausible but sometimes false output, calls for a renewed emphasis on designing for trust. We need to design systems and train users so that they develop appropriate levels of trust with a system, and so that we can avoid well known biases such as automation bias. Users needed to be helped to develop their critical thinking skills and not fall into complacency. Intelligibility in general helps here, but more specific design considerations, mostly drawn from [11] and [18], include:

- Take account of users who may vary in expertise, AI literacy, and task familiarity in designing the user interface and feedback from the system.

- Identify ways to assess automation bias from telemetry and monitor with changes to the interface. Use onboarding techniques and tutorials to make users aware that overreliance is a common phenomenon, giving them examples of correct and incorrect output. It is especially crucial to set the appropriate levels of trust in early use of

- a system. Be transparent about a system's strengths and limitations, as well as intended uses.

- Endow systems with well-calibrated confidences in their output and guide output based on these confidences, including the provision of indications of uncertainty in a system's output.

## 4 Conclusion

The rapid evolution and widespread adoption of generative AI technologies have profoundly influenced various domains, from daily communication and workflows, to providing administrative and decision support in specialized fields like medicine and law. As AI becomes more commonplace in our lives, our perception and language surrounding it evolve. Drawing parallels with aviation, the "AI co-pilot" metaphor captures important aspects of the relationship between AI and humans, where the machine assists but does not dominate. The human remains in control, making critical decisions, while the AI offers support, expertise, and backup. Lessons from aviation highlight the potential pitfalls of over-relying on automation, which brings several recognized challenges, including diminished vigilance, challenges in transferring control, de-skilling, and misplaced trust. To foster a harmonious and productive human-AI partnership, it is an imperative to prioritize human agency, keep humans actively engaged, provide opportunities to enhance human skills, design more intelligible AI systems, and employ methods that cultivate an appropriate level of trust in AI output and guidance. By integrating insights from both Human-Computer Interaction and Human Factors Engineering and investing more intensively in the rising subdiscipline of Human-AI Interaction & Collaboration, we can navigate this transformative era, ensuring that AI serves as a valuable co-pilot, enhancing human capabilities while respecting our agency and expertise.